\documentclass[12pt,preprint]{emulateapj}

\usepackage{natbib}

\shorttitle{Determination of transport coefficients by coronal seismology}
\shortauthors{Wang \& Ofman}
\newcommand{\adv}{    {AdSpR}}
\newcommand{\ag}{    {AnGeo}}
\newcommand{\natp}{    {NatPh}}
\newcommand{\newa}{    {NewA}}

\begin{document}

\title{Determination of transport coefficients by coronal seismology of flare-induced slow-mode waves: Numerical parametric study of 1D loop model}

\email{tongjiang.wang@nasa.gov}

\author{Tongjiang Wang}
\affil{Department of Physics, Catholic University of America,\\
   620 Michigan Avenue NE, Washington, DC 20064, USA}
\affil{NASA Goddard Space Flight Center, Code 671, \\
   Greenbelt, MD 20770, USA}

\author{Leon Ofman}
\affiliation{Department of Physics, Catholic University of America,\\
   620 Michigan Avenue NE, Washington, DC 20064, USA}
\affiliation{NASA Goddard Space Flight Center, Code 671, \\
   Greenbelt, MD 20770, USA}
\affiliation{Visiting, Tel Aviv University, Israel}

\begin{abstract}
Recent studies of a flaring loop oscillation event on 2013 December 28 observed by the Atmospheric Imaging Assembly (AIA) of the {\it Solar Dynamics Observatory} ({\it SDO}) have revealed the suppression of thermal conduction and significant enhancement of compressive viscosity in hot ($\sim$10 MK) plasma. In this study we aim at developing a new coronal seismology method for determining the transport coefficients based on a parametric study of wave properties using a 1D nonlinear MHD loop model in combination with the linear theory. The simulations suggest a two-step scheme: we first determine the effective thermal conduction coefficient from the observed phase shift between temperature and density perturbations as this physical parameter is insensitive to the unknown viscosity; then from the loop model with the obtained thermal conduction coefficient, we determine the effective viscosity coefficient from the observed decay time using the parametric modeling. With this new seismology technique we are able to quantify the suppression of thermal conductivity by a factor of about 3 and the enhancement of viscosity coefficient by a factor of 10 in the studied flaring loop. Using the loop model with these refined transport coefficients, we study the excitation of slow magnetoacoustic waves by launching a flow pulse from one footpoint. The simulation can self-consistently produce the fundamental standing wave on a timescale in agreement with the observation.
\end{abstract}


\section{Introduction}
Slow magnetoacoustic oscillations of hot coronal loops were first discovered with the high-resolution imaging spectrometer, SOHO/SUMER, in flare emission lines (mainly Fe\,{\sc xix} and Fe\,{\sc xxi} as periodic variations of the Doppler shift \citep[e.g.,][]{wan02, wan03a}. Similar Doppler shift oscillations were also detected in the flare lines, S\,{\sc xv} and Ca\,{\sc xix}, with Yohkoh/BCS \citep{mar05, mar06}. These oscillations are characterized by relatively long periods on the order of a few minutes to a few tens of minutes and a rapid decay with a ratio of the decay time to the oscillation period of about unity. They are mainly interpreted as the fundamental standing slow waves because their periods correspond to twice the acoustic travel time along the loop and there is a quarter-period phase lag between velocity and intensity disturbances detected in some cases \citep[see][for a review]{wan11}. In addition, standing and (upwardly) propagating slow waves were also detected in warm ($\sim1-2$ MK) coronal loops with Hinode/EIS in emission lines Fe\,{\sc xi}--Fe\,{\sc xv} exhibiting Doppler shift and intensity (density) oscillations \citep{mar08,erdt08,wan09a,mar10,van11}. These oscillations are characterized with small amplitudes (a few percentage of the sound speed) and very weak damping.

Recently, longitudinal intensity oscillations in flare loops were detected with the {\it Solar Dynamics Observatory (SDO)}/Atmospheric Imaging Assembly (AIA) in the 94 \AA\ and 131 \AA\ channels \citep[e.g.,][]{kum13}. These oscillations, shown with the observed properties (such as oscillation periods and decay times) matching the SUMER oscillations, have been interpreted as either a standing slow-mode wave \citep{wan15, nis17} or a reflecting propagating slow wave \citep{kum13, kum15, man16}. It is well known that whether a compressive slow wave is a standing or propagating wave can be definitely identified based on the phase relationship between density and velocity perturbations: an inphase relation indicates a propagating wave \citep[e.g.][]{wan09a, wan09b}, whereas a quarter-period phase relation indicates a standing wave \citep[e.g.][]{sak02, wan03b}. Although not obvious as the phase-lag criteria, it is possible to identify the wave mode (standing or propagating) solely based on its signatures in intensity as suggested by theoretical and forward modeling studies \citep{nak04, tsi04, tar07, tar08, yuan15, fan15, wan18}. In the spatial manner the fundamental standing wave shows the anti-phase oscillations between the two legs \citep[e.g.,][]{yuan15}, whereas the reflecting propagating wave for a single pulse exhibits a ``zigzag'' pattern in a time-distance plot made along the loop and the propagating speeds are close to the speed of the sound in the flaring loop \citep[e.g.,][]{wan18}. In the temporal behavior the standing wave of the fundamental mode (or a harmonic) follows a sinusoidal function (in the linear uniform loop case; it is assumed that gravitational stratification of the density can be neglected in a hot flaring loop since the gravitational scale height is larger than the loop height) whereas the reflecting wave pulse does not. The AIA observations show that in most cases the detrended light curve of disturbances is characterized by a decaying sine function instead of a few quasi-periodic pulsations \citep[e.g.,][]{kum13,kum15,wan15}. This may suggest that the waves are dominated by a single harmonic component in power, or the waves could be interpreted as the standing wave of a dominant harmonic. Like the SUMER oscillation events \citep[e.g.][]{wan05}, the initiation of the AIA longitudinal waves is also associated with small flares (GOES B or C-class) at the loop footpoint \citep{kum13,kum15,wan15,nis17}. Magnetic extrapolation and emission features suggest that the confined flares are produced by reconnections at a coronal null point in a fan-spine magnetic topology \citep{wan15,wan18,kum15}. The impulsive energy release triggers longitudinal oscillations in the hot (outer) spine loop, which may be heated by energetic particles or heat flux from the reconnection region \citep{sun13, wan15}. Exceptionally, a standing slow-mode wave event was observed in ($\sim$1 MK) fanlike coronal loop system triggered by a global EUV wave originating remotely \citep{pan17}. 

\begin{figure*}
\epsscale{1.0}
\plotone{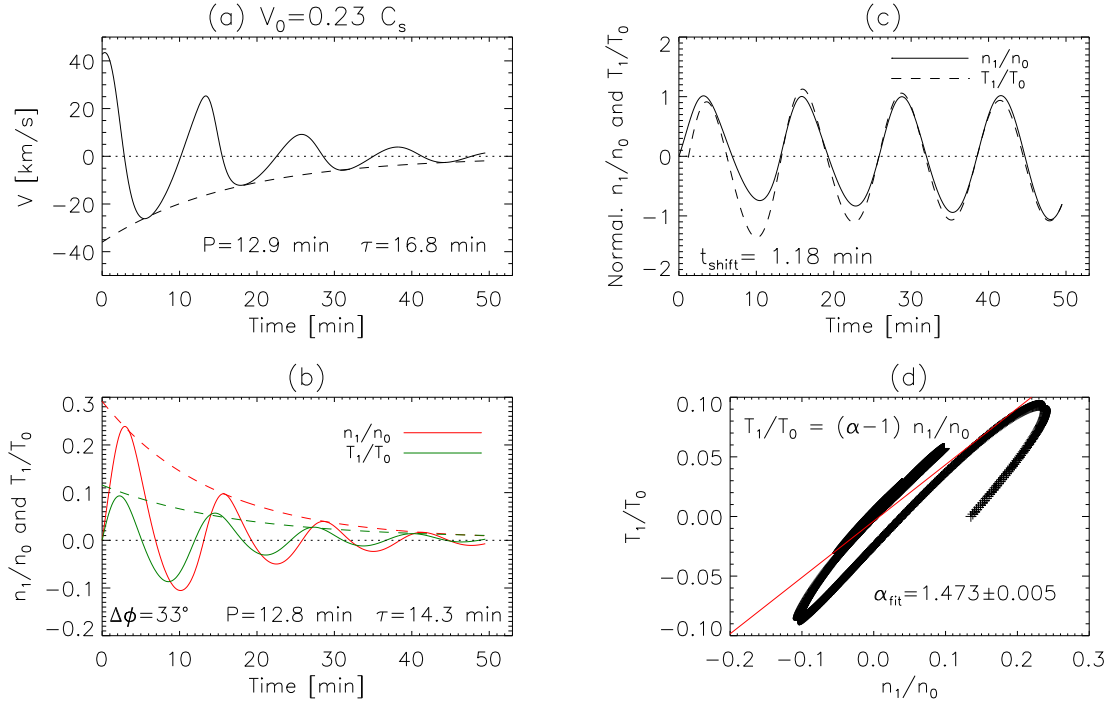} 
\caption{ \label{fgmdl} Simulations of the fundamental standing slow-mode wave with the classical thermal conduction and compressive viscosity coefficients. (a) Temporal evolution of the velocity $V$ at the location $x=158$ Mm. (b) Temporal evolution of the perturbed density $n_1/n_0$ and the perturbed temperature $T_1/T_0$ at the same location. The exponential decay time fit follows the dashed line. The oscillation period ($P$) and decay time ($\tau$) marked in (b) are measured from the density perturbations, and the phase shift ($\Delta{\phi}$) between $n_1$ and $T_1$ is measured by cross correlation. (c) Amplitude-normalized density and temperature perturbations. The temperature profile has been corrected relative to the density profile by a phase shift of $t_{\rm shift}$=1.18 minutes, which corresponds to the maximum correlation between them. (d) Scatter plot of the perturbed density and temperature (pluses) and its best fit (solid line). The measured polytropic index ($\alpha_{\rm fit}$) is marked on the plot. }
\end{figure*}

A large number of numerical and theoretical studies have been dedicated to understanding rapid damping of the slow magnetoacoustic waves observed in hot coronal loops. \citet{ofm02} were the first to develop nonlinear 1D MHD model of the damping of moderately nonlinear slow magnetoacoustic waves in hot coronal loops that included the effects of thermal conduction, viscosity, and self-consistent heating. Based on the results of the modeling guided by SOHO/SUMER observations, \citet{ofm02} first suggested that thermal conduction is the dominant damping mechanism of the observed slow magnetoacoustic waves. Further investigations include the various effects such as compressive viscosity \citep{dem03, sig07, wan18}, radiative cooling \citep{pan06, sig07, pro18}, gravitational stratification \citep{men04, sig07}, non-equilibrium ionization \citep{bra08}, shock dissipation \citep{ver08}, nonlinearity \citep{rud13,afa15,nak19}, geometry \citep{ogr07, ogr09, sel09}, temperature inhomogeneity \citep{erd08, abe12}, cooling background \citep{alg13, alg14, bah18}, and the wave-caused misbalance between heating and cooling processes \citep{kum16, nak17, kol19}. Much attention has been also paid to the excitation mechanism of standing slow-mode waves. The 3D isothermal MHD simulations show that a global standing slow wave can be excited by a velocity pulse or an impulsive onset of flows near one footpoint \citep{sel09, ofm12}. The 3D ideal MHD simulations show that the global standing slow wave can also be excited by a fast-mode wave with an impact angle almost parallel to the loop’s plane \citep{pas09}. Theoretical analysis and simulations of 1D loop models show that a footpoint heating pulse with a duration much shorter than the wave period generates only (reflected) propagating slow waves \citep{sel05, tar05, tar08}. This conclusion was confirmed by \citet{fan15} using a 2.5D MHD model including effects of the chromosphere. \citet{wan18} found that the anomalously enhanced viscosity can explain the quick formation of standing slow waves seen in observations. In addition, numerical simulations by \citet{men06} showed that intermittent patterns of the standing slow wave can be produced by random energy releases near either one or both footpoints of the loop, likely by interference. This provides a possible excitation mechanism for weakly-damped slow standing waves observed with Hinode/EIS in warm coronal loops.

Coronal seismology is a diagnostic tool for determining fundamental plasma parameters in the solar atmosphere by comparing the model-predicted behavior of MHD waves with observations \citep{zai75, zai82, rob83, nak05}. Reviews of recent progress in this topic can be found in \citet{liu14}, and \citet{wan16}. With this technique the observations of slow magnetoacoustic waves have been used to derive some important physical parameters in coronal loops, such as the magnetic field strength \citep{wan07, jes15}, longitudinal structuring \citep{mce06,sri10,mac10,sri13}, polytropic index \citep{van11,wan15,kri18}, viscosity coefficient \citep{wan15}, and heating function \citep{tar07, rea16, rea18, kol19}. 

By analyzing a longitudinal oscillation event on 2013 December 28 in AR 11936 observed by SDO/AIA, \citet{wan15} found that the temperature and density disturbances are nearly in-phase and the measured polytropic index is close to 5/3. This suggests the strong suppression of thermal conduction in hot flaring coronal loops. Based on the linear wave theory, they predicted that the classical thermal conductivity is suppressed by at least a factor of 3 and the classical compressive viscosity coefficient is enhanced by a factor of 15 at most. \citet{wan18} validated this          result using a 1D nonlinear MHD model with a flow driver at one footpoint. Their simulations showed that the model with fully-suppressed (or no) thermal conduction and 15 times enhanced viscosity can reproduce many observed wave properties better than the model with the classical transport coefficients, such as the timescale of standing wave formation, the phase relationship between density and temperature perturbations, the polytropic index, and the characteristics of linear wave (manifested by the time profiles of the observed density and temperature perturbations following well a damped-sine function). In this study, we extend the coronal seismology technique based on the linear wave theory to that based on a parametric study of wave properties using nonlinear numerical modeling that enables us to determine both effective thermal conduction and viscosity coefficients more accurately. We describe the loop model in Section~\ref{sctmdl}. We present the results of parametric study in Section~\ref{sctpsp}, their comparison with the analytical solutions in Section~\ref{sctcas}, and the method for determination of transport coefficients in Section~\ref{sctdtc}. We model the formation of the standing wave using the seismology-determined transport coefficients in Section~\ref{sctmwe}. The discussion and conclusions are given in Section~\ref{sctdac}. 

\begin{figure*}
\epsscale{0.9}
\plotone{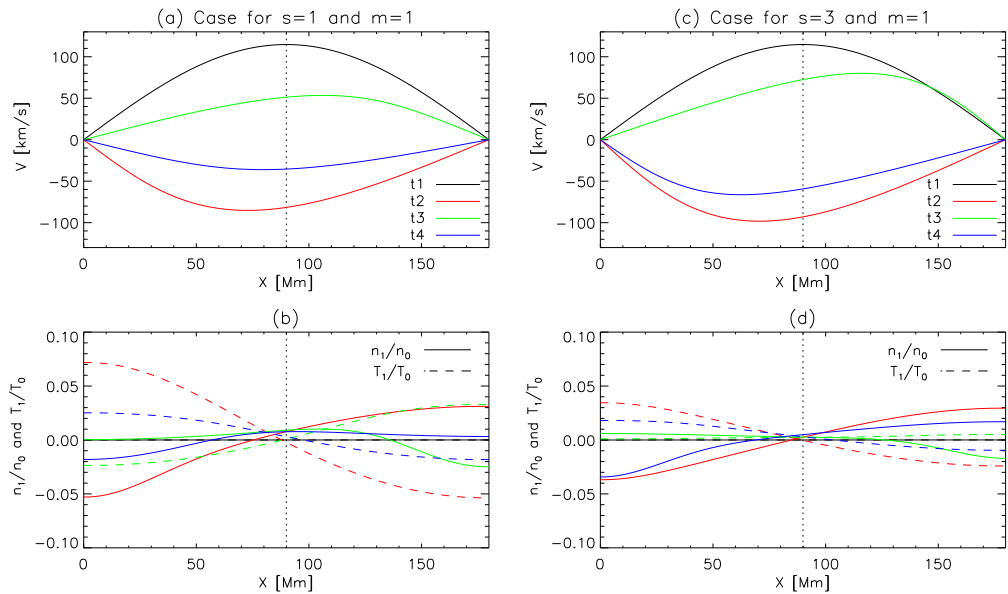}
\caption{ \label{fgnln} Spatial distributions of (a) the velocity, and (b) the perturbed density (solid line) and perturbed temperature (dashed line) along the loop at $t=$ 0.0, 6.4, 12.8, 19.2 minutes (marked $t_1-t_4$ corresponding to 0, $P/2$, $P$, $3P/2$ where $P$ is the wave period) for the case with the classical thermal conduction and classical viscosity coefficients. (c) and(d) Same as (a) and(b) but the spatial profiles at $t=$ 0.0, 6.0, 12.0, and 18.0 minutes (corresponding to 0, $P/2$, $P$, $3P/2$) for the case with the 3 times reduced thermal conduction coefficient and the classical viscosity coefficient (i.e. taking $s$=3 and $m$=1). A vertical dotted line indicates the middle position of the loop. The accompanying animation shows the evolution of the velocity, density, and temperature perturbations in the two cases from $t$=0.0 minutes to $t$=47.6 minutes. The animation duration is 17 s. (An animation of this figure is available.)}
\end{figure*}

\section{Loop model}
\label{sctmdl}
The damping of a standing slow-mode wave is simulated by setting a velocity profile of the fundamental mode on an initially uniform coronal loop in a one-dimensional Cartesian geometry. The spatial profile of the velocity wave is given by
\begin{equation} 
V(x, t=0)=V_0\,{\rm sin}\left(\frac{k\pi{x}}{L}\right),
\end{equation}
where $V_0$ is the wave amplitude at $t=0$, and $k$ is the mode number ($k=1$ for the fundamental mode),
$x$ is the spatial coordinate along the magnetic field and $L$ is the total length of the loop. The plasma motion along the loop is governed by the following set of nonlinear hydrodynamic equations: 
\begin{eqnarray}
 \frac{{\partial}{\rho}}{\partial{t}} +\frac{\partial}{\partial{x}} (\rho{V})&=&0, \label{eqct}\\
 \rho\left(\frac{{\partial}{V}}{\partial{t}}+V\frac{\partial{V}}{\partial{x}} \right)&=&-\frac{{\partial}{p}}{\partial{x}}+ F_{\nu}, \label{eqmm}\\
 \frac{{\partial}{T}}{\partial{t}}+(\gamma-1)T\frac{\partial{V}}{\partial{x}}+V\frac{\partial{T}}{\partial{x}}&=&\frac{(\gamma-1)m_p}{2k_B}\left(\frac{1}{\rho}\right)(S_{\nu}+H_c), \label{eqeg}
\end{eqnarray} 
where $\rho$ is the mass density, $p=2\rho{k_B}T/m_p$ the gas pressure, $T$ the temperature, $m_p$ the proton mass, $k_B$ the Boltzmann constant, and $\gamma$=5/3 in the corona. The equations include the terms for compressive viscosity and thermal conduction, but neglect the gravity and radiative losses. The assumption is justified for hot flaring loops at $T\simeq10$ MK since the gravitational scale height is larger than the height of a typical hot flaring loop \citep[see][]{wan15, wan18}. The viscous force ($F_{\nu}$) and the viscous heating ($S_{\nu}$) due to compressive viscosity are expressed in terms of
\begin{eqnarray}
F_{\nu}&=&\frac{4}{3}\eta_0\left(\frac{\partial^2V}{\partial{x}^2}\right), \\
S_{\nu}&=&\frac{4}{3}\eta_0\left(\frac{\partial{V}}{\partial{x}}\right)^2,
\end{eqnarray}
with the classical Braginskii compressive viscosity coefficient for ions (considering protons only), 
\begin{equation}
\eta_0=2.23\times10^{-15}\,\frac{T^{5/2}}{{\rm ln}\Lambda} ~~~ {\rm g~cm}^{-1}\,{\rm s}^{-1},
\end{equation}
where ${\rm ln}\Lambda=8.7-{\rm ln}(n^{1/2}T^{-3/2})$ is the Coulomb logarithm, weakly dependent on $T$ and the number density $n$. For this study, ${\rm ln}\Lambda\approx{22}$, so $\eta_0=10^{-16}T^{5/2}$ \citep[see][]{hol86}. The heat conduction term along the magnetic field is 
\begin{equation}
H_c=\frac{\partial}{\partial{x}}\left(\kappa_{0}T^{5/2}\frac{\partial{T}}{\partial{x}}\right), 
\end{equation}
where $\kappa_0=7.8\times10^{-7}~{\rm erg~cm}^{-1}\,{\rm s}^{-1}\,{\rm K}^{-7/2}$ is the classical Spitzer thermal conduction coefficient for electrons \citep{spi56, spi62}. In this parametric study we assume the initial density ($n_0$) and temperature ($T_0$) along the loop to be uniform, and use $n_0=2.6\times10^9$ cm$^{-3}$, $T_0=9$ MK and $L=180$ Mm measured from observations \citep[see][]{wan15}. For these loop parameters we calculate the adiabatic sound speed $C_s=(\gamma{k_B}T_0/\mu{m_p})^{1/2}=166(T_0/{\rm MK})^{1/2}=498$ km~s$^{-1}$ (here taking $\mu=0.5$ is to be consistent with that used in Equations~(\ref{eqct})--(\ref{eqeg}) for simulations in Section~\ref{sctpsp}), the classical compressive viscosity coefficient $\eta_0=24.6~{\rm g~cm}^{-1}\,s^{-1}$, and the classical thermal conductivity parallel to the magnetic field $\kappa_\|=\kappa_{0}T_0^{5/2}=1.90\times10^{11}~{\rm erg~cm}^{-1}\,{\rm s}^{-1}\,{\rm K}^{-1}$. Using the same method as in \citet{wan18}, we take the initial velocity amplitude $V_0=(n_m/n_0)C_s$=115 km~s$^{-1}$, where $n_m/n_0=0.23$ is the maximum amplitude of density perturbations measured from observations.

The MHD equations are solved by adopting the fourth-order Runge-Kutta method for integration in time and fourth-order spatial derivatives using 256 grid points \citep{ofm02, wan18}. Numerical convergence is tested by doubling the resolution and comparing the results. The boundary conditions at both ends of the loop are $V(0, t)=V(L, t)=0$ and zero-order extrapolation for the rest of the variables.

\begin{figure*}
\epsscale{1.0}
\plotone{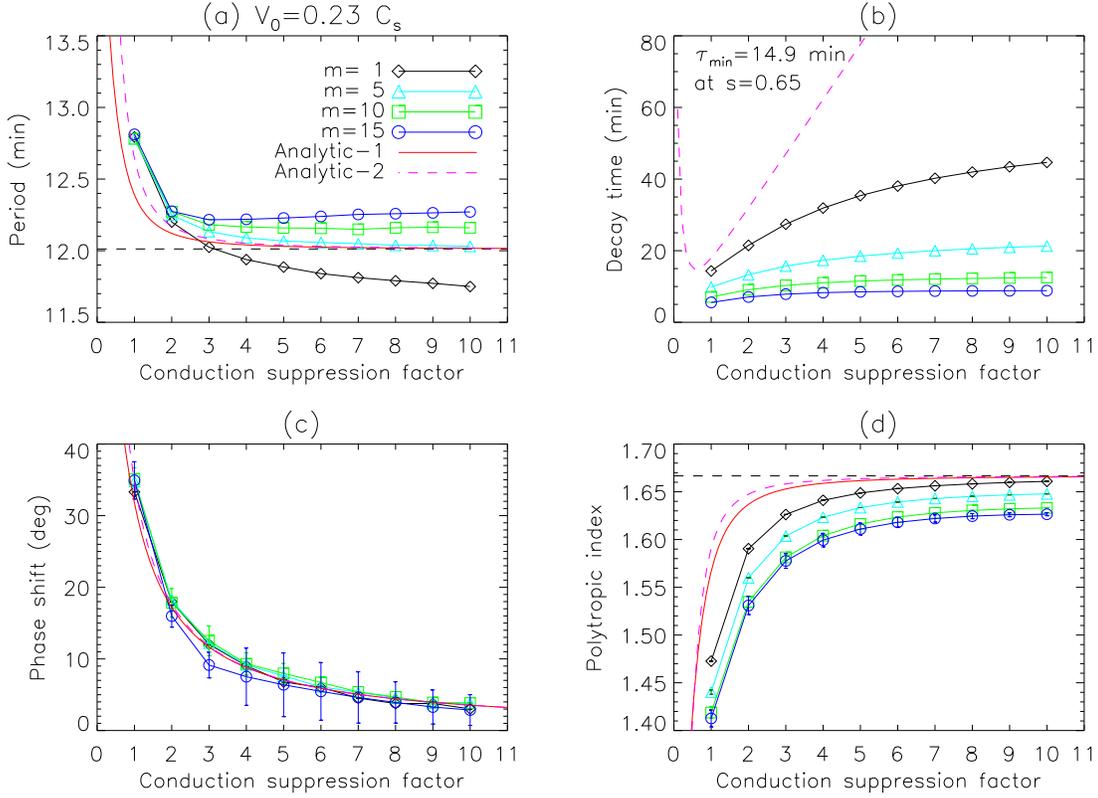}
\caption{ \label{fgpsc} Dependence of wave properties on the thermal conduction suppression factor ($s$) for different viscosity enhancement factors ($m$) based on parametric simulations of the fundamental slow wave with initial velocity amplitude $V_0=0.23\,C_s$. (a) Wave period. (b) Decay time. (c) Phase shift between the density and temperature perturbations. (d) Polytropic index. The red solid lines in (a), (c), and (d) represent the analytical solutions named Analytic-1 (Equations~(\ref{eqphs}), (\ref{eqaps}) and (\ref{eqpds})), whereas the purple dashed lines represent the analytical solutions named Analytic-2, where the curves in (a) and (b) are calculated from the dispersion relation (Equation~(\ref{eqdsp})) and the curves in (c) and (d) are calculated with Equations~(\ref{eqphg}) and (\ref{eqapg}). The horizontal dashed line in (a) indicates $P_0=2L/C_s$, and the horizontal dashed line in (d) indicates $\gamma=5/3$. }
\end{figure*}

\begin{figure*}
\epsscale{1.0}
\plotone{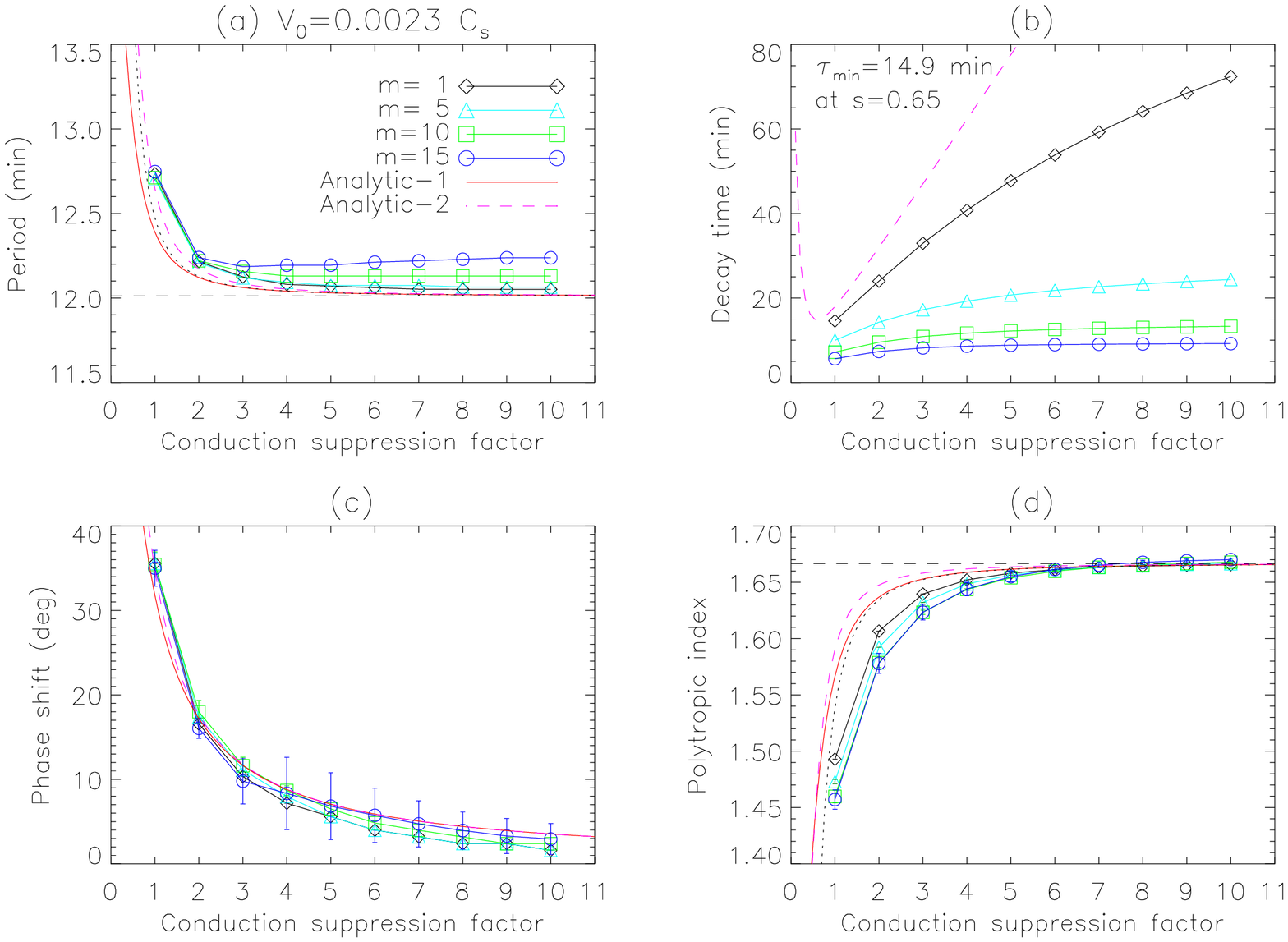} 
\caption{ \label{fgsac} Same as Figure~\ref{fgpsc} but for the case with initial velocity amplitude 100 times smaller as a control experiment. The black dotted lines in (a) and (d) represent the Taylor approximation (Equations~(\ref{eqap}) and (\ref{eqpp})) of the Analytic-1 solutions.}
\end{figure*}

\section{Parametric study of wave properties}
\label{sctpsp}
Using the 1D loop model described above, we study the dependence of wave properties of the fundamental mode on different transport coefficients by parametric simulations. The purpose is to find a reliable method for determining the transport coefficients from observables such as wave period, decay time, phase shift, and polytropic index. The advantage of using the parametric simulation is twofold. First, it allows to account for the combined effects of thermal conduction and compressive viscosity on the wave evolution (particularly, when their dissipation rates are comparable). Second, it allows to account for the nonlinearity effect on the wave properties when an initial perturbation has a large amplitude compared to the phase speed ($C_s$). Motivated by the finding of \citet{wan15}, we define the thermal conduction suppression factor ($s$) and the compressive viscosity enhancement factor ($m$) as follows:
\begin{equation}
 s=\frac{\kappa_0}{\kappa_s}, ~~~~
 m=\frac{\eta_m}{\eta_0},
\end{equation}
where $\kappa_s$ ($\le\kappa_0$) is the suppressed thermal conduction coefficient and $\eta_m$ ($\ge\eta_0$) is the enhanced viscosity coefficient. We perform the parametric simulations using the loop model with $\kappa_s$ ($s=1-10$) and $\eta_m$ ($m=1-18$). 

We use the case for $s=1$ and $m=1$ (i.e., the model with the classical thermal conduction and viscosity coefficients) to demonstrate the simulation results and the methods used to measure wave properties (see Figure~\ref{fgmdl}). Figures~\ref{fgmdl}(a) and (b) show the temporal evolution of perturbed velocity ($V$), density ($n_1/n_0$), and temperature ($T_1/T_0$) measured at a location ($x=0.88L$) near the loop footpoint. We estimate the wave period ($P$) by averaging time intervals between successive peaks in the velocity (or density) profile, and estimate the decay time ($\tau$) by fitting the wave peaks to an exponentially-damped function ($f(t)=A_0 + A_1\,e^{-t/\tau}$). The measured $P$ and $\tau$ are marked on the plots. The simulations show the presence of a phase shift between density and temperature perturbations as predicted from linear MHD theory \citep[see Section~\ref{sctcas} and also][]{owe09, van11, wan15, wan18}. To measure the phase shift ($\Delta{\phi}$), we first normalize the damped oscillations using the method $(s(t)-s_0)/(f_s(t)-s_0)$, where $s$ represents $n_1/n_0$ or $T_1/T_0$, $f_s(t)$ is the best-fit exponentially-damped function, and $s_0$ is the background calculated by averaging $s(t)$ over time. We then apply the cross correlation to the normalized time profiles and obtain a time shift $t_{\rm shift}$=1.18 minutes between $n_1/n_0$ and $T_1/T_0$ (see Figure~\ref{fgmdl}(c)). The phase shift is calculated as $\Delta{\phi}=360{\degr}(t_{\rm shift}/P)=33{\degr}$, where $P$=12.8 minutes is the wave period measured for $n_1/n_0$. We measure the polytropic index $\alpha$ by fitting the scaling between $T_1/T_0$ and $n_1/n_0$ (after first removing their phase shift $\Delta{\phi}$) to the following equation \citep[see][]{wan15, wan18, kri18},
\begin{equation}
  \frac{T_1}{T_0}=(\alpha-1)\frac{n_1}{n_0}.  \label{eqalp}
\end{equation} 
Using this method we obtain $\alpha_{\rm fit}=1.473\pm0.005$ (see Figure~\ref{fgmdl}(d)). 

Figure~\ref{fgnln} demonstrates the evolution of the perturbed velocity, density, and temperature profiles along the loop in two cases: one with $s=1$ and $m=1$ (Panels (a) and (b)) and the other with $s=3$ and $m=1$ (Panels (c) and (d)). An animation of Figure~\ref{fgnln} is available in the online version. One can clearly see the wave steepening in the velocity profile caused by the generation of higher harmonics due to the nonlinearity. The resultant behavior of a periodic shifting of the peak position in amplitudes along the loop from the middle (where the peak position is expected in the linear case for the fundamental mode) is consistent with the prediction by nonlinear wave theory \citep{rud13, kum16}. \citet{rud13} showed that for typical SUMER oscillations the nonlinearity effect on the damping time becomes important when the Mach number of the initial velocity perturbation is of the order of or larger than 0.2. For the wave event modeled here, it was estimated that $V_0/C_s\approx0.23$ from the AIA observations \citep{wan15,wan18}, belonging to such a case. This also suggests that taking into account the nonlinearity effect will help improving the seismological accuracy (e.g. by using nonlinear MHD modeling).

\begin{figure*}
\epsscale{1.0}
\plotone{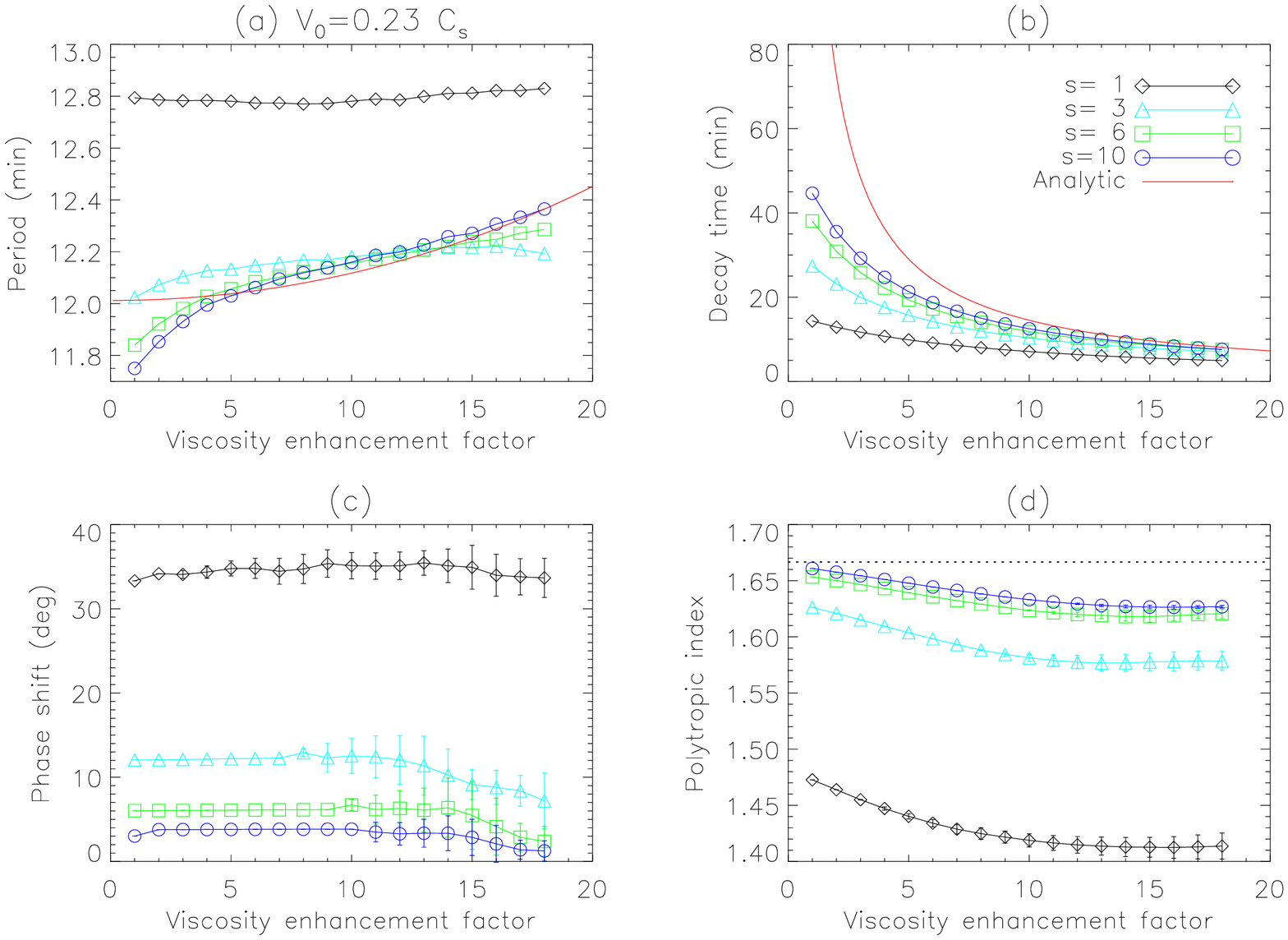}
\caption{ \label{fgpsv} Dependence of wave properties on the viscosity enhancement factor ($m$) for different thermal conduction suppression factors ($s$) based on parametric simulations of the fundamental slow wave with initial velocity amplitude $V_0=0.23\,C_s$. (a) Wave period. (b) Decay time. (c) Phase shift between the density and temperature perturbations. (d) Polytropic index. The red solid lines in (a) and (b) represent the solutions of linear theory (see Equations~(\ref{eqvp}) and (\ref{eqvtd})). The horizontal dotted line in (d) indicates $\gamma=5/3$. }
\end{figure*}

\subsection{Dependence on thermal conduction suppression}
\label{sstdtc}
We first examine the dependence of measured wave properties on the thermal conduction suppression factor in different cases with the enhanced viscosity factor $m=1$, 5, 10, 15 (see Figure~\ref{fgpsc}).  Figure~\ref{fgpsc}(a) shows that the wave period decreases rapidly with the suppression of thermal conduction when the suppression factor is small ($s\la3$), but the curve becomes flat for larger $s$, especially in the case with the enhanced viscosity. In addition, the dependence of wave period on $s$ is clearly different for the cases with different $m$ when the suppression factor is large ($s\ga3$). The features suggest that the effective thermal conduction coefficient may be determined from the observed wave period only when the conduction suppression is small (e.g., $s<2$). In addition, this method requires the accurate measurement of the 3D loop length used in the model, which is generally difficult for coronal observations with present observations. Figure~\ref{fgpsc}(b) shows that the decay time increases with the suppression of thermal conduction, but the increase becomes slower for the larger viscosity. We estimate an increase of decay time from $s=1$ to 10 by a factor of 3.1, 2.2, 1.8, 1.6  for m=1, 5, 10, 15, respectively. In addition, the dependence of decay time on $s$ is clearly different for the cases with different $m$. Since the decay time depends on both $s$ and $m$, the effective thermal conduction coefficient cannot be uniquely determined from the observed decay time. 

Figure~\ref{fgpsc}(c) shows that the phase shift between density and temperature perturbations is nearly independent on the viscosity. This independence of viscosity suggests that the phase shift is an ideal observable to use for diagnosing the thermal conduction suppression. Note the larger uncertainty in measured phase shifts for the viscosity enhancement case with $m=15$. This is because the damping time of the waves is very short (compared to the period) for the present condition (see Figure~\ref{fgpsc}(b)). It does not make sense to cross-correlate the normalize temporal profiles of density and temperature oscillations over the whole range ($t=0-50$ minutes) as tiny amplitudes are practically undetectable. Thus we limit the measurement to ranges varying from $t=0-4\tau$ to $t=0-6\tau$, and take the average over 11 measurements. The error bars are the standard deviation, reflecting the uncertainty due to the range selection for cross correlation. Measurements of the polytropic index and its error bar are obtained in the same way. For instance, we obtain $\Delta{\phi}= 6.4\pm4.4{\degr}$ and $\alpha_{\rm fit}=1.611\pm0.006$ in the case of $s=5$ and $m=15$. Figure~\ref{fgpsc}(d) shows a clear dependence of polytropic index on the suppression of thermal conduction when the suppression factor is small ($s\la4$), but the curve becomes flat when $s$ is larger. In addition, the dependence of polytropic index on $s$ is slightly different for the cases with different $m$. This difference will cause the uncertainty in determination of the effective thermal conduction coefficient from the observed polytropic index, and the uncertainty increases with the polytropic index (becoming large especially when $\alpha\ga1.6$) since the curves become flatter.

To check the effect of nonlinearity on the result, we repeat the above parametric simulations with the initial perturbation amplitude 100 times smaller, i.e., taking $V_0=0.0023\,C_s$ as a control experiment. Figures~\ref{fgsac}(a) and (b) show the dependence of wave period and decay time on the conduction suppression factor, indicating that the main disparity between the small- and large-amplitude cases is in the behavior of the curve for $m=1$. In the small perturbation condition, the curve of wave period for $m=1$ is much closer to the cases with the large viscosity enhancement, whereas the curve of decay time for $m=1$ is much more deviated from the cases with larger $m$. These differences can be explained based on the study of \citet{wan18} who found that the large viscosity enhancement greatly reduces the effects of nonlinearity. The steepened wave front caused by nonlinearity in the large-amplitude condition propagates at a speed faster than the linear wave. This implies that in the small-amplitude case the wave period for $m=1$ becomes longer and so its curve becomes closer to those curves with larger $m$ which are less sensitive to changes of the initial velocity amplitude as their nonlinearity is strongly suppressed by large viscosity enhancement. For the same reason, the difference in the curves of decay time for $m=5$, 10 and 15 between the small- and large-amplitude cases is also small. In the small-amplitude case the slope of the decay time curve for $m=1$ is larger than that in the large-amplitude case. This feature can be explained by the enhancement of damping rates by nonlinearity for large-amplitude waves \citep{ver08,rud13}, whose effect becomes more evident when the thermal conduction suppression is large. We estimate that the decay time from $s=1$ to 10 in the small-amplitude case increases by a factor of 5.0, 2.4, 1.9, 1.6 for m=1, 5, 10, 15, respectively.

Figures~\ref{fgsac}(c) and (d) show the dependence of phase shift and polytropic index on the conduction suppression factor, respectively, in the small-amplitude case. Their behaviors are similar to those in the large-amplitude case. Since they are insensitive to the variation of viscosity, both the phase shift and the polytropic index can be used to determine the effective thermal conduction coefficient in the case of small (or linear) perturbations. Note that the uncertainty of the derived conduction coefficient is large when the observed phase shift is less than about 10${\degr}$ or the observed polytropic index is larger than about 1.65 as the curves are flatter in such conditions.

\begin{figure*}
\epsscale{1.0}
\plotone{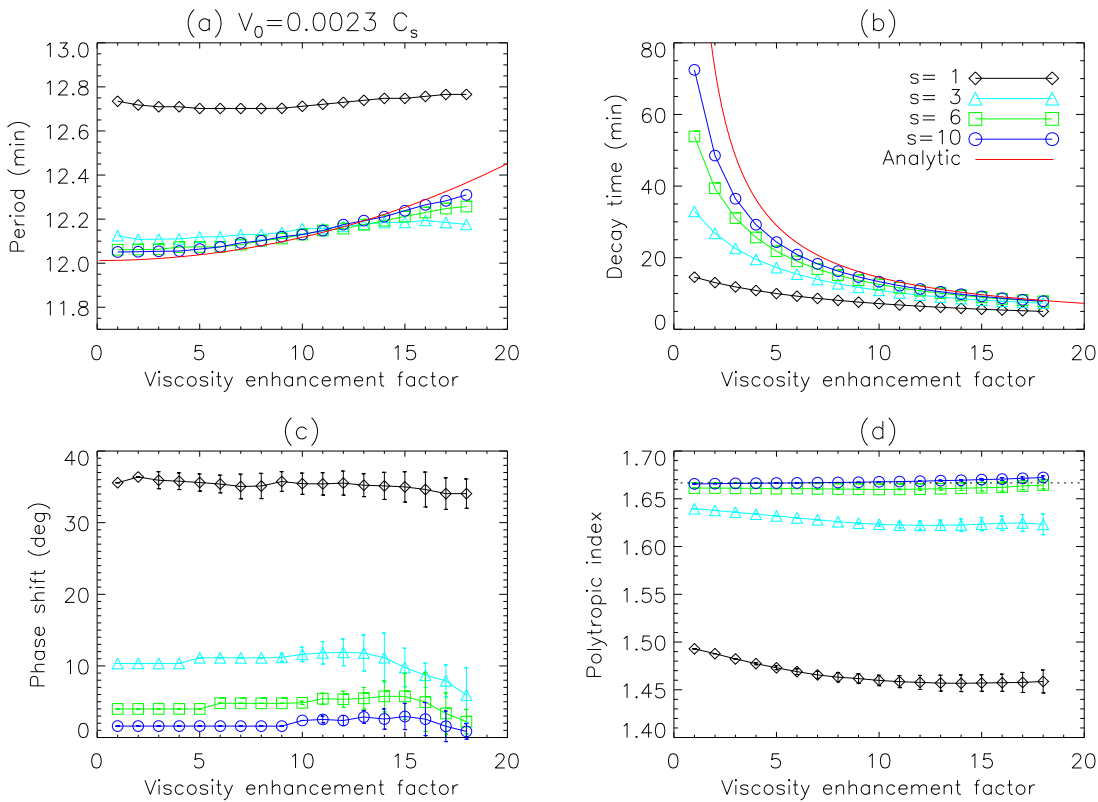}
\caption{ \label{fgsav} Same as Figure~\ref{fgpsv} but for the case with initial velocity amplitude 100 times smaller as a control experiment.}
\end{figure*}

\subsection{Dependence on compressive viscosity enhancement}
\label{sstdcv}
We now examine the dependence of measured wave properties on the viscosity enhancement factor in different cases with the conduction suppression factor $s=1$, 3, 6, 10 for initial velocity $V_0=0.23\,C_s$ (see Figure~\ref{fgpsv}). Figure~\ref{fgpsv}(a) shows that the wave period is nearly independent on the viscosity enhancement when $s=1$, while the period slightly increases with the viscosity (especially when $m\la5$) when the thermal conduction is highly suppressed (when $s\geqslant3$). Figure~\ref{fgpsv}(b) shows that the decay time decreases drastically with the enhancement of viscosity, especially when $m$ is relatively small ($m\la10$). The curves of decay time with different thermal conduction factors are distinctly different. We estimate that the decay time decreases from $m=1$ to 18 by a factor of 2.9, 3.9, 5.0, 5.9 for $s=1$, 3, 6, 10, respectively. The prominent dependence of decay time on the viscosity suggests that the effective viscosity coefficient may be determined from the observed decay time once the effective thermal conduction coefficient is obtained (e.g., from the observed phase shift or polytropic index as mentioned in Section~\ref{sstdtc}). Figure~\ref{fgpsv}(c) shows that the phase shift is nearly independent on the viscosity enhancement except for the condition of $m\ga14$, where the phase shift shows a weak decrease. It also shows that the error bars in measurements increase with the enhancement of viscosity. Figure~\ref{fgpsv}(d) shows that the polytropic index is weakly dependent on the viscosity enhancement, indicating a slow decrease when $m\la12$. In addition, the curves of polytropic index for the cases of large thermal conduction suppression ($s=6$ and 10) are very close.  

Figure~\ref{fgsav} shows the dependence of the wave properties on the viscosity enhancement in the condition of small initial velocity amplitude ($V_0=0.0023\,C_s$) to check the importance of the nonlinear effect. We find that compared to the large-amplitude case the dependence of wave period on the viscosity enhancement becomes even weaker for the curves with $s=3$, 6, 10, showing much slower increase rates when $m\la5$ (Figure~\ref{fgsav}(a)). From Figure~\ref{fgsav}(b) we estimate that the decay time decreases from $m=1$ to 18 by a factor of 2.9, 4.5, 6.9, 9.2 for $s=1$, 3, 6, 10, respectively. It shows that the decrease rates of decay time for the cases with $s\geqslant3$ are larger than those in the large-amplitude case, indicating that the nonlinear effects caused by large-amplitude perturbations become prominent when the thermal conduction is largely suppressed (see also Figure~\ref{fgnln}(c)). The dependence of decay time on the viscosity enhancement factor for the case of $s=1$ is almost the same (with difference $\la2\%$ on average) between the small- and large-amplitude cases. This feature suggests that the wave damping by large thermal conductivity in hot plasma significantly dominates over the damping by nonlinearity at the large amplitude ($V_0/C_s=0.23$) in the loop model with the classical thermal conduction coefficient \citep[see also discussions in Section 5 of][]{wan18}. In addition, the comparison between the small- and large-amplitude cases indicates that the the effect of nonlinearity on decay times becomes very weak when the viscosity enhancement is larger (with $m\ga10$).

Figure~\ref{fgsav}(c) shows that the phase shift in the small-amplitude case is nearly independent on the viscosity enhancement, similar to the large-amplitude case. Likewise, the phase shift for the cases with $s=3$, 6, 10 also shows a weak decrease as in the large-amplitude case for larger viscosity enhancements ($m\ga14$). This trend is caused by a systematic error in normalizing the damped oscillations of density and temperature perturbations when we measure the phase shift. We find that for the large-$m$ cases the fitting error to the exponentially-damped function $f(t)$ (especially in temperature) increases with the viscosity enhancement. Figure~\ref{fgsav}(d) shows that the polytropic index for the cases with $s=1$, 3 shows a slow decrease with the viscosity enhancement, similar to the large-amplitude case, while it shows almost no changes for the cases with $s=6$, 10 in the small-amplitude cases. This weak dependence is related to a systematic error caused by the data selection criterion we used in measurements of the phase shift (see Section~\ref{sstdtc}), where since we discarded the data with very small amplitudes for the large-$m$ cases it led to the best fits for the polytropic index towards the scattered data with larger amplitudes and so causing a reduction in the fitted slope (see Figure~\ref{fgmdl}(d)). 

\section{Comparison with linear analytical solutions}
\label{sctcas}
To better understand the results of the parametric study described above, we compare them with the analytical solutions derived based on the linear MHD theory.  

Let us first consider the case when only thermal conduction is present. It is well known that dissipation of the slow wave by thermal conduction leads to the presence of a phase shift ($\Delta{\phi}$) between temperature and density perturbations which can be expressed in the following form on the assumption of the phase speed $V_p=\omega/k\approx C_s$ and weak damping ($\tau/P\gg1$) \citep[see][]{van11,wan15}:
\begin{eqnarray}
  {\rm tan}\,\Delta{\phi}&=& 2\pi\gamma d,      \label{eqphi} \\
  (\gamma-1) {\rm cos}\,\Delta{\phi}&=& \alpha-1, \label{eqpa}
\end{eqnarray}
where $\alpha$ is the polytropic index as defined in Equation~(\ref{eqalp}), and $d$ is the thermal ratio \citep[see][]{dem03}, defined as 
\begin{equation}
d=\frac{(\gamma-1)\kappa_\|T_0 \rho_0}{{\gamma}^2p_0^2 P}\approx 4.1\left(\frac{T_0^{3/2}}{n_0 P}\right),
\label{eqd}
\end{equation}
where $\rho_0=n_0 m_p$, $p_0=2n_0k_BT_0$, and $P=2\pi/\omega$ is the wave period. By replacing $d$ in Equation~(\ref{eqphi}) with $d_s=d/s$, we obtain the predicted phase shift and polytropic index as a function of the thermal conduction suppression factor:
\begin{eqnarray}
  \Delta{\phi}_s&=& {\rm tan}^{-1}\frac{2\pi\gamma{d}}{s},    \label{eqphs} \\
  \alpha_s &=& \frac{\gamma-1}{\sqrt{(2\pi\gamma{d}/s)^2+1}}+1.     \label{eqaps}   
\end{eqnarray}
Considering the wave speed for the polytropic process \citep[see][]{wan18},
\begin{equation}
  V_p=\left(\frac{\alpha}{\gamma}\right)^{1/2}C_s,  \label{eqpvp}
\end{equation}
we predict the wave period using $P=2L/V_p$, in combination with Equations~(\ref{eqaps}) and (\ref{eqpvp}) giving
\begin{equation}
 P_s=\left(\frac{2L}{C_s}\right)\left(\frac{\gamma}{\frac{\gamma-1}{\sqrt{(2\pi\gamma{d}/s)^2+1}}+1}\right)^{1/2}.   \label{eqpds}
\end{equation}
To see more obviously the dependence of wave period and polytropic index on thermal conduction suppression factor, we apply the Taylor expansion considering that $\Delta=(2\pi\gamma{d}/s)^2\ll{1}$. For example, with the observed physical parameters $T_0=9$ MK, $n_0=2.6\times10^9$ cm$^{-3}$ and $P=12$ minutes, we have $d=0.06$ and $\Delta=0.1$ for $s=2$. Equations~(\ref{eqaps}) and (\ref{eqpds}) are then approximated to first order as
\begin{eqnarray}
  \alpha_s &\approx& \gamma\left(1-\frac{2C}{s^2}\right), \label{eqap} \\
  P_s  &\approx& P_0\left(1+\frac{C}{s^2}\right),   \label{eqpp} 
\end{eqnarray}
where $P_0=2L/C_s$ and $C=(\gamma-1)\gamma\pi^2{d^2}$.

We show the above analytical results for wave period, phase shift and polytropic index (in red solid lines) in Figures~\ref{fgpsc} and ~\ref{fgsac}. We find a good agreement between the analytical and numerical results of the phase shift in both the small- or large-amplitude perturbation cases (see Panel (c)). The analytical solutions of wave period and polytropic index approach to their adiabatic limits rapidly with the increase of the thermal conduction suppression factor (see Panels (a) and (d)). Figures~\ref{fgsac} (a) and (d) show that they are closely approximated by their first-order Taylor expansion (in dotted lines) for $s\ga1$. It is obvious that the numerical results of wave period and polytropic index more close to their analytical solutions in the case of small-amplitude perturbations.

\begin{figure*}
\epsscale{1.0}
\plotone{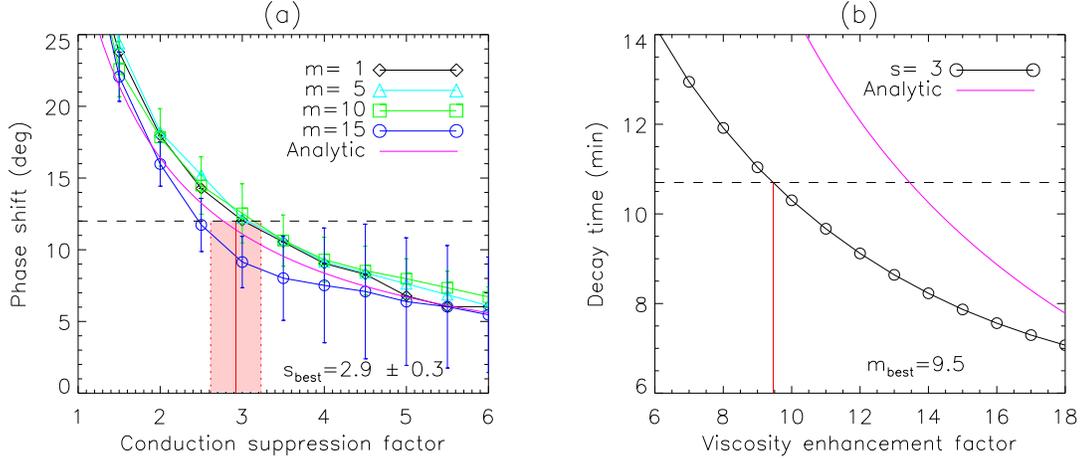}
\caption{ \label{fgsch} Determination of transport coefficients using the parametric modeling scheme. (a) Dependence of the phase shift on the conduction suppression factor ($s$). The horizontal dashed line indicates the phase shift ($\Delta\phi_{\rm obs}$) measured between the density and temperature wave signals from the SDO/AIA observations. The vertical red line represents the mean ($s_{\rm best}$) of $s$-factors corresponding to $\Delta\phi_{\rm obs}$, obtained from the curves for the viscosity enhancement factor $m$=1, 5, 10, and 15. The pink band shows the $\pm{1}\sigma$ standard deviation. (b) Dependence of the decay time on the viscosity enhancement factor. The horizontal dashed line indicates the decay time ($\tau_{\rm obs}$) measured from the density oscillation of the SDO/AIA observations. The vertical red line represents the obtained viscosity enhancement factor ($m_{\rm best}$) corresponding to $\tau_{\rm obs}$ from the curve for $s$=3 ($\approx{s}_{\rm best}$). The pink solid curves in both plots are the solution of linear theory, which are same as in Figure~\ref{fgpsc}(c) and Figure~\ref{fgpsv}(b)).}
\end{figure*}

Given the more general condition that the wave frequency is complex ($\omega=\omega_r + i\omega_i$) and the phase speed $V_p=\omega_r/k$, we can derive the improved equations for the phase shift and polytropic index \citep[see][]{wan18}:
\begin{eqnarray}
\Delta{\phi_s}&=& {\rm tan}^{-1}\frac{2\pi\gamma\left(\frac{d}{s}\right)\left(\frac{C_s}{V_p}\right)^2{\rm cos}\,\psi/\sqrt{1+\chi^2}}{1-2\pi\gamma\left(\frac{d}{s}\right)\left(\frac{C_s}{V_p}\right)^2{\rm sin}\,\psi/\sqrt{1+\chi^2}},     \label{eqphg} \\
\alpha_s &=&\frac{(\gamma-1){\rm cos}\,\Delta{\phi}}{1-2\pi\gamma\left(\frac{d}{s}\right)\left(\frac{C_s}{V_p}\right)^2{\rm sin\,\psi}/\sqrt{1+\chi^2}}+1, \label{eqapg}
\end{eqnarray}
where
\begin{equation}
d=\frac{(\gamma-1)\kappa_\|T_0 \rho_0}{(2\pi/\omega_r){\gamma}^2p_0^2},  \label{eqdg}
\end{equation}
and $\chi=\omega_i/\omega_r$ and $\psi={\rm tan}^{-1}\chi$. Note that here the wave frequency $\omega$ is a function of the thermal conduction suppression factor $s$. We calculate the values of $\omega_r$ and $\omega_i$ from the dispersion relation for the fundamental standing wave with a wavelength of $2L$. The dispersion relation can be derived under the assumption of all disturbances in the form $e^{i(\omega{t}-kx)}$ as \citep[see][]{dem03}:
\begin{equation}
 P_0^3\omega^3-i(4\pi^2\gamma{d_0}/s)P_0^2\omega^2-4\pi^2{P_0}\omega+i(16\pi^4{d_0}/s)=0, \label{eqdsp}
\end{equation}
where $P_0=2L/C_s$=12 minutes and $d_0=(\gamma-1)\kappa_\|T_0\rho_0/(\gamma^2 p_0^2 P_0)$=0.06. For each value of $s$, we find the roots of $\omega$ using the IDL function, FZ\_ROOTS, which is based on the Laguerre's method. We then calculate the phase shift and polytropic index using Equations~(\ref{eqphg}) and (\ref{eqapg}). Panels (a) and (b) of Figures~\ref{fgpsc} and~\ref{fgsac} show the calculated wave period ($P=2\pi/\omega_r$) and decay time ($\tau=1/\omega_i$) in purple dashed lines, respectively. We find that the analytical solution of the wave period, calculated from the dispersion relation (named ``Analytic-2" method for convenience), agrees with the simulated result for $m=1$ in the small perturbation case better than the analytical solution using Equation~(\ref{eqpds}) (named ``Analytic-1" method) (see Figure~\ref{fgsac}(a)). Figure~\ref{fgsac}(b) shows that there is a minimum decay time ($\tau_{\rm min}=14.9$ minutes at $s_{\rm min}=0.65$). This feature is well known \citep[e.g.][]{dem03}, and the result ($d_{\rm min}=d_0/s_{\rm min}\approx0.1$) is consistent with the theoretical prediction that a minimum damping time occurs when thermal ratio $d$=0.1 due to thermal conduction alone. The analytical solution of decay time is more closed to the simulated decay times for $m=1$ in the small-amplitude case than in the large-amplitude case. Figures~\ref{fgsac}(c) and (d) show that the disparity between the solutions of Analytic-1 and Analytic-2 is very small for both phase shift and polytropic index, justifying the approximation of $V_p\approx C_s$ and weak damping in this loop model.

We now consider the case when the slow waves are damped by compressive viscosity alone. By substituting the velocity perturbation, $v=v_0\,e^{i(\omega{t}-kx)}$, into the velocity wave equation \citep[see][]{sig07}, the dispersion relation for the fundamental mode ($k=\pi/L$) can be derived:
\begin{equation}
  \omega^2-i\left(\frac{16\pi^2 m\varepsilon_0}{3P_0}\right)\omega-\left(\frac{4\pi^2}{P_0^2}\right)=0,   \label{eqvdp}
\end{equation}
where $m$ is the viscosity enhancement factor, $P_0=2L/C_s$, and the dimensionless ratio $\varepsilon_0=\eta_0/(\rho_0{C_s}^2P_0)=3.15\times10^{-3}$. We can solve Equation~(\ref{eqvdp}) for the solution of the complex frequency ($\omega=\omega_r+i \omega_i)$, and thus obtain the wave period ($P=2\pi/\omega_r$) and decay time ($\tau=1/\omega_i$) as
\begin{eqnarray}
   P_m   &=&P_0\left(1-\frac{16\pi^2\varepsilon_0^2 m^2}{9}\right)^{-1/2}, \label{eqvp} \\
  \tau_m &=&P_0\left(\frac{3}{8\pi^2\varepsilon_0}\right)\frac{1}{m}.  \label{eqvtd}
\end{eqnarray}
Since the term $\Delta=16\pi^2\varepsilon_0^2 m^2/9\ll{1}$ (e.g., $\Delta=0.07$ for $m=20$), Equation~(\ref{eqvp}) can be approximated to first order as
\begin{equation}
 P_m\approx{P_0}\left[1+\left(\frac{8\pi^2\varepsilon_0^2}{9}\right)m^2\right]. \label{eqvpp}
\end{equation}
This indicates that the wave period increases almost quadratically with the enhancement of compressive viscosity. In addition, Equation~(\ref{eqvtd}) indicates that the decay time is inversely proportional to the viscosity enhancement. We show the theory-predicted wave period (Panel (a)) and decay time (Panel (b)) in red solid lines in Figures~\ref{fgpsv} and~\ref{fgsav}. We find that the analytical solution of wave period is roughly consistent with the simulated results with strong suppression of thermal conduction. The agreement is better especially for the case with $s=10$ and when the viscosity enhancement factor is larger ($m\ga5$), confirming the significant role of viscosity in suppressing the nonlinearity. In addition, their agreement is better in the small-amplitude case than in the large-amplitude case (Figure~\ref{fgpsv}(a) and Figure~\ref{fgsav}(a)). The simulated results of decay time for the cases with $s=6$, 10 are also more closed to its analytical solution in the small-amplitude case than in the large-amplitude case (Figure~\ref{fgpsv}(b) and Figure~\ref{fgsav}(b)). From the linearized HD equations of mass continuity and energy with an equation of state, we can derive the relation $T_1/T_0=(\gamma-1)n_1/n_0$. This implies that the temperature and density perturbations are in phase and the polytropic index is equal to the adiabatic index of 5/3 in the linear theory including compressive viscosity only. Thus some small variations of the simulated phase shift and polytropic index with the viscosity enhancement factor (e.g., in the case with $s=10$; see Figures~\ref{fgpsv}(c) and (d)) are likely due to a limited thermal conduction, nonlinear effect (including the viscous heating), and some systematic errors with the used methods in measurements (see Section~\ref{sstdcv}).

\section{Scheme for determination of transport coefficients}
\label{sctdtc}
Based on the above analysis of parametric simulations in combination with linear theories, we propose the following procedure to determine the effective transport coefficients from observations using 1D nonlinear HD modeling:

Step~1. We estimate the thermal conduction suppression factor from the observed phase shift between temperature and density oscillations based on its dependence on the simulated phase shift (see Figure~\ref{fgsch}(a)). For $\Delta\phi_{\rm obs}\approx12\degr$ as measured in \citet{wan15}, we obtain $s=3.0$, 3.1, 3.1, 2.5 by interpolation from the simulated data for $m=1$, 5, 10, 15, respectively. The average gives $s_{\rm best}=2.9\pm0.3$, indicating that the effective thermal conductivity is smaller than the classical value by a factor of about three. 

Step~2. We estimate the viscosity enhancement factor from the observed decay time for density oscillations based on its dependence on the simulated decay time for the case with $s=s_{\rm best}$ as determined in Step~1 (see Figure~\ref{fgsch}(b)). For $\tau_{\rm obs}=10.7\pm4.0$ minutes as measured in \citet{wan15}, we obtain $m_{\rm best}=9.5\pm7.0$ from the simulated data for $s=3$. This result suggests that the compressive viscosity is enhanced by almost one order of magnitude compared to the classical value in this flaring loop.

\begin{figure*}
\epsscale{1.0}
\plotone{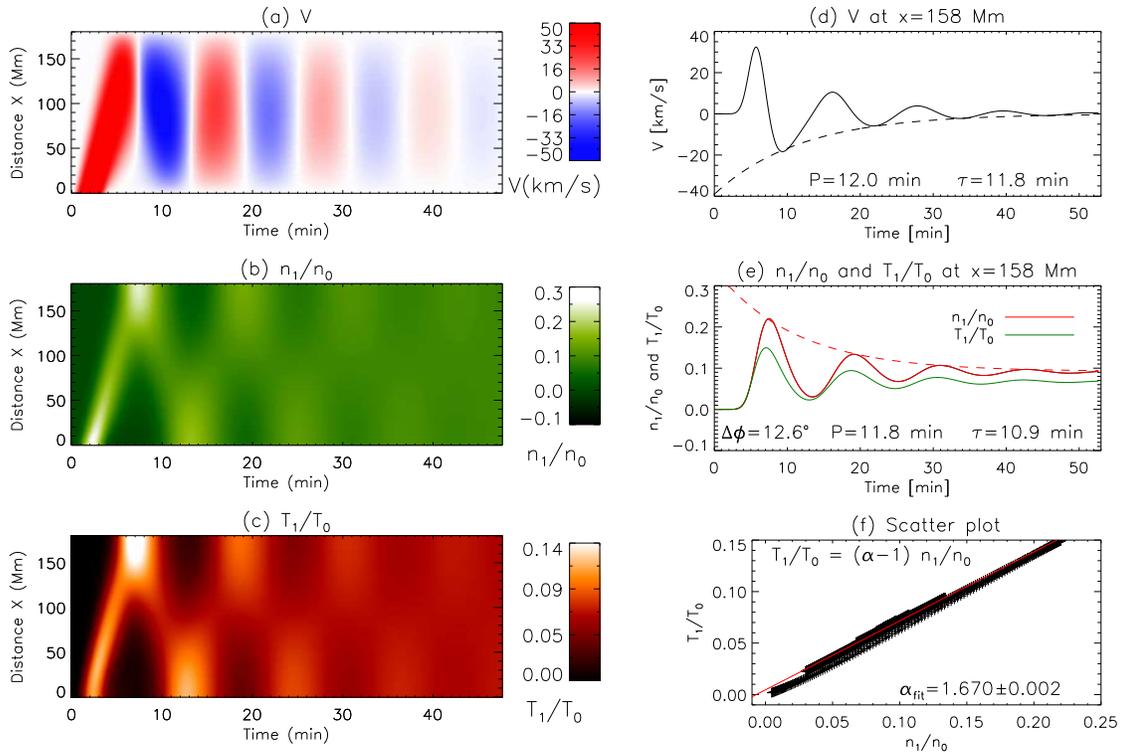}
\caption{ \label{fgexc} Simulations of wave excitation by an initial flow pulse at $x$=0 using the loop model with the seismology-determined thermal conduction and viscosity coefficients (taking $s$=3 and $m$=9.5 obtained from Figure~\ref{fgsch}). (a)-(c) Time distance maps for velocity ($V$), perturbed density ($n_1/n_0$), and perturbed temperature ($T_1/T_0$) along the loop. Temporal profiles of (d) $V$, and (e) $n_1/n_0$ and $T_1/T_0$ at the location $x=158$ Mm. The dashed line represents the exponential decay time fit. (f) The scatter plot (pluses) and its best fit (solid line). In (d)-(f) the measured oscillation period ($P$), decay time ($\tau$), phase shift ($\Delta{\phi}$) between $n_1$ and $T_1$, and polytropic index ($\alpha_{\rm fit}$) are marked on the plots.}
\end{figure*}

\section{Modeling the formation of the standing wave}
\label{sctmwe}
As in \citet{wan18}, we simulate the wave excitation using a flow pulse driven at one footpoint of the loop. The loop model is same as described in Section~\ref{sctmdl} but with the seismology-determined transport coefficients, i.e., the effective thermal conduction coefficient $\kappa_{\rm cs}=\kappa_0/s_{\rm best}$ and the effective viscosity coefficient $\eta_{\rm cs}=\eta_0 m_{\rm best}$ as obtained in Section~\ref{sctdtc}.

To simulate the flare-induced perturbation, we inject an impulsive flow along the magnetic field at one end,
\begin{equation}
V(x=0, t)  =   \left\{
\begin{array}{ll}
\frac{1}{2}V_0 \left[1-{\rm cos}\left(\frac{2\pi t}{t_{\rm dur}}\right)\right] &  \quad (0\leqslant t \leqslant t_{\rm dur}),\\ 
 0 &  \quad (t > t_{\rm dur}). \label{eqdrv}
\end{array}\right.,
\end{equation}
where $V_0$ is the pulse amplitude, set to be $V_0=0.23C_s=115$ km~s$^{-1}$, and $t_{\rm dur}$ is the pulse duration, taken as 4 minutes as estimated from observations \citep[see][]{wan18}. The boundary conditions at both ends of the loop are $V(0, t)$=$V(L, t)$=0 except the flow injection $V(0, 0\leqslant t\leqslant t_{\rm dur}$).

We show the temporal evolution of velocities, perturbed densities and temperatures along the loop in Figures~\ref{fgexc}(a)-(c). It is shown that a fundamental standing slow wave pattern forms right after the initial propagating disturbance is reflected at the remote footpoint ($x=L$). This is indicated by some theory-predicted characteristics such as in-phase velocity oscillations along the loop, antiphase density (or temperature) oscillations between the two legs, and a quarter-period phase shift between velocity and density perturbations \citep[e.g.][]{tar08, yuan15}. This simulation with the seismology-determined transport coefficients self-consistently explains the observed quick setup of the fundamental standing wave in a flaring loop \citep{wan15}. For quantitative comparison with the observed wave properties we analyze the temporal evolution of velocity, density, and temperature perturbations at a location ($x=0.88L$) near the remote footpoint of the loop (see Figures~\ref{fgexc}(d)-(f)). We measure the wave period, decay time, phase shift between density and temperature perturbations, and polytropic index using the same methods as described in Section~\ref{sctpsp}. We obtain $P=11.8$ minutes, $\tau=10.9$ minutes, $\Delta\phi=12.6\degr$, and $\alpha=1.670\pm0.002$. The results agree well with the measurements from SDO/AIA observations in \citet{wan15} giving $P_{\rm obs}=12.4\pm1.0$ minutes and $\tau_{\rm obs}=10.7\pm4.0$ minutes for density oscillations, $\Delta\phi_{\rm obs}\approx12\degr$, and $\alpha_{\rm obs}=1.64\pm0.08$. Thus, good agreements between the simulation and observation in the timescale of the fundamental standing wave formation and wave properties validate our determined transport coefficients and the new coronal seismology method.

\section{Discussion and conclusions}
\label{sctdac}
We have performed a numerical parametric study about the fundamental standing slow wave using a 1D loop model, motivated by recent studies of a flaring loop oscillation event on 2013 December 28 in AR 11936 observed with SDO/AIA \citep{wan15, wan18}. The dependence of wave properties (such as the period, decay time, phase shift between temperature and density perturbations, and polytropic index) on transport coefficients (thermal conduction and compressive viscosity) is examined based on 1D nonlinear HD simulations in combination with linear analytical solutions. The simulations show that the dependence of the phase shift on the thermal conduction suppression factor is insensitive to the viscosity and nonlinearity, whereas the dependence of the decay time on the viscosity enhancement factor is most obvious compared to the other wave properties. These features allow us to develop a new coronal seismology technique determining the transport coefficients with a two-step scheme: We first estimate the effective thermal conduction coefficient from the observed phase shift using loop models with different assumed viscosity coefficients, and then we estimate the effective viscosity coefficient from the observed decay time using the parametric modeling with the conduction coefficient obtained in the first step. With this technique we find that the classical thermal conductivity is suppressed by a factor of about 3 and the classical viscosity coefficient is enhanced by a factor of 10 in this hot flaring loop.

Note that the results in our study are obtained from the loop model with uniform equilibrium temperature and density. If assuming an inhomogeneous loop model that contains the chromosphere and transition region and the initial loop atmosphere in hydrostatic equilibrium \citep[e.g.][]{tar08, bra08}, both the equilibrium temperature and density will vary along the loop. Considering the profiles of relatively lower temperature and higher density near the loop's footpoints, the averaged equilibrium temperature and density over the entire loop will be slightly lower and higher, respectively, compared to those used in the uniform loop model, which were measured from the coronal part of the loop \citep[see][]{wan15}. From Equations~(\ref{eqphi}) and~(\ref{eqd}) we expect that the phase shift simulated in the inhomogeneous loop model would be somewhat smaller than in the uniform loop model. This implies that the thermal conduction suppression factor obtained from the uniform loop assumption may be slightly overestimated. This small effect of the present simplifying assumption needs to be verified in more detail in the future study. In addition, we have assumed the loop model initially in equilibrium, whereas hot flaring loops typically experience a cooling process during the decay phase \citep[e.g.][]{sun13, wan15}. This implies that the phase shift between temperature and density fluctuations should evolve with time. If such changes in phase shift are detectable, we may derive temporal evolution of the effective transport coefficients by applying our suggested seismology method to a quasi-static loop model when the cooling timescale is much longer than the wave period. Such measurements of the transport coefficients may help to elucidate the long lasting cooling phase in some flares \citep[e.g.][]{woo11, hoc12, liu13, liu15, sun13, li14, dai18a, dai18b, zho19}, when they are used as the more strict constraints on loop heating models.

In our loop model, only two dominant wave dissipation mechanisms (thermal conduction and compressive viscosity) are included. Neglecting the role of other damping mechanisms thus implies that the obtained viscosity enhancement factor in this study is an upper limit. For example, \citet{sig07} showed that in stratified loops nonlinear viscous dissipation causes a reduction of decay time by about 3--13\% compared to the unstratified loops. \citet{bra08} found that the radiative loss resulted from a non-equilibrium ionization reduces the damping time by up to $\sim$10\% in comparison to the equilibrium case. \citet{ver08} showed that shock dissipation at large amplitudes can damp the oscillation at a rate comparable to that by thermal conduction alone. However, our simulations show that the effect of nonlinearity on damping is significantly weakened in the presence of the anomalously enhanced viscosity.  In addition, \citet{pan06} showed that optically thin radiation plays a neglecting role in strong-damped ($\tau/P\sim1$) oscillations (which our studied case belongs to), while it may increase a dissipation by up to 20\% apart from thermal conductivity and viscosity in weak-damped ($\tau/P\geqslant2$) oscillations, which are, however, uncommon in observations of hot loop oscillations \citep[see][]{wan03a,wan07}.

Using the loop model with the obtained transport coefficients we have simulated the excitation of slow magnetoacoustic waves by launching a flow pulse at one footpoint of the loop. We find that the fundamental standing wave pattern forms immediately after the initial propagating perturbation is reflected at the other footpoint on a timescale in agreement with the observation. This simulation result confirms the wave excitation mechanism proposed in \citet{wan18}, where simulations of higher harmonic waves reveal that the decay time is proportional to the period squared in the case with the enhanced viscosity by a factor of 15 and no thermal conduction. Such a relation suggests that the large viscosity facilitates the dissipation of higher harmonic components in the initial pulse, so favoring the quick setup of the fundamental standing mode. In contrast, in the other case with the classical thermal conduction and viscosity coefficients, the scaling relation between the decay time and the period is linear, and a timescale of several reflections is required for the initial propagating disturbance to form the standing wave pattern. This latter loop model may explain the reflected propagating feature of a longitudinal wave event reported in \citet{kum13}. 

By analyzing the same event as in \citet{kum13} using the Fourier cosine transform, \citet{nak19} showed the decay of different spatial harmonics in a similar behavior and suggested that the expected frequency-dependent dissipation is compensated by nonlinearity. However, the interpretation of their result needs to be validated in new studies.

\acknowledgments
The work of TW and LO was supported by NASA grants 80NSSC18K1131 and the NASA Cooperative Agreement NNG11PL10A to CUA. The work of TW was also supported by the NASA grant 80NSSC18K0668.

\end{document}